# Embedding Knowledge Graph of Patent Metadata to Measure Knowledge Proximity

Guangtong Li✉ , L Siddharth, Jianxi Luo

*Data-Driven Innovation Lab, Engineering Product Development Pillar,*

*Singapore University of Technology and Design, 8 Somapah Road, 487372, Singapore.*

---

## Abstract

Knowledge proximity refers to the strength of association between any two entities in a structural form that embodies certain aspects of a knowledge base. In this work, we operationalize knowledge proximity within the context of the US Patent Database (knowledge base) using a knowledge graph (structural form) named 'PatNet' built using patent metadata, including citations, inventors, assignees, and domain classifications. We train various graph embedding models using PatNet to obtain the embeddings of entities and relations. The cosine similarity between the corresponding (or transformed) embeddings of entities denotes the knowledge proximity between these. We compare the embedding models in terms of their performances in predicting target entities and explaining domain expansion profiles of inventors and assignees. We then apply the embeddings of the best-preferred model to associate homogeneous (e.g., patent-patent) and heterogeneous (e.g., inventor-assignee) pairs of entities.

---

✉ Corresponding author: guangtong_li@mymail.sutd.edu.sg

# 1. Introduction

The constituents of a knowledge base could assume various structural forms (e.g., citation network) depending on the relations (e.g., cited by) intended to be captured by such forms. The entities that constitute such structural forms could be associated using *a posteriori* measure of 'proximity' that quantifies the strength of structural relations. Such a measure of proximity could be termed 'knowledge proximity' when it is derived from a structural form that embodies certain aspects of a knowledge base.

In this article, we operationalize knowledge proximity within the US patent database, wherein, the primary entity patent is linked to other entities such as inventors (e.g., 4074775 – "Dawn Tan"), assignees (e.g., 336083 – "Microsoft Corporation"), domain – subsection (e.g., H01 – "Basic Electric Elements"), and domain – group (e.g., H01L – "Semiconductor Devices") and various other patents through citations. Scholars have proposed quantitative measures for knowledge proximity that have often been utilized to associate homogenous pairs (e.g., inventor-to-inventor) of entities. Such measures also denote the opposite of "technological distance" or "knowledge distance" (Leydesdorff et al., 2014).

While various proximity measures have been utilized to associate pairs of domains (Alstott et al., 2017b; Yan & Luo, 2017a), e.g., through co-citation for demonstrating analogical transfer of concepts (Luo et al., 2021), such measures are less applicable to heterogeneous pairs (e.g., inventor-domain) as these measures only capture limited aspects of the patent database through individual relations such as <patent, cite, patent>, <assignee, *own*, patent>, <domain, *contain*, patent>, etc. Our research, therefore, recognizes the need for a structural form that embodies all types of entities (e.g., patent, assignee, domain) and relations (e.g., citation, ownership, classification).

As explained in Section 3.1, we capture the links among entities in the US Patent Database (1976-2020) using a flexible structural form – a knowledge graph that captures entities and relations as a set of facts {<head entity, relation, tail entity>}. We capture the following relations: <patent, *cite*, patent>, <inventor, *write*, patent>, <assignee, *own*, patent>, <group, *contain*, patent> and <subsection, *comprise*, groups>. We apply various embedding algorithms (Section 3.2) to the knowledge graph thus constructed to obtain embeddings of entities and relations. The cosine similarity between corresponding (or transformed) embeddings of entities denotes the knowledge proximity. In Section 4, we evaluate these embedding algorithms in terms of predicting target entities ($\langle ?, r, t \rangle$ or $\langle h, r, ? \rangle$) and explaining the domain expansion history of all inventors and assignees. In Section 5, we apply knowledge proximity to associate different pairs of entities and make inferences therefrom.

# 2. Related Work

In this section, we review the existing patent-based knowledge proximity measures (Section 2.1) and generic approaches to embedding knowledge graphs (Section 2.2).

## 2.1. Knowledge Proximity Measures

Leydesdorff and Vaughan (2006) propose knowledge proximity between a **pair of patents** as the number of forward and backward citations shared by these. Aharonson and Schilling (2016) vectorize patents using 9,864 classification digits and propose knowledge proximity as the Euclidean distance. Often applied to patent documents (Feng, 2020; Whalen et al., 2020), Latent Semantic Analysis (LSA) involves Singular Value Decomposition (SVD) performed on a term-document matrix for obtaining document vectors (Deerwester et al., 1990). Scholars have proposed knowledge proximity as the cosine similarity between patent vectors obtained through LSA (An et al., 2021; Gerken & Moehrle, 2012; Yoon & Kim, 2012).

Diestre and Rajagopalan (2012) measure knowledge proximity between a **pair of assignees** (e.g., Merck and Pfizer) as the number of overlapping patent classes jointly owned by them. Scholars have obtained vector representations of assignees using the distribution of patents across domains. Using such vector representations, they measure knowledge proximity between assignees as cosine, Pearson's Correlation Coefficient (Guan & Yan, 2016), and Euclidean distance (Ahuja, 2000).

Scholars have adopted the co-occurrence of patents (Dibiaggio & Nesta, 2005; Teece et al., 1994) or classification codes (Schoen et al., 2012) as knowledge proximity between a **pair of domains**. They propose vector space representation of domains (e.g., "Alloys," C22C) using the distribution of citations across all domains (Kay et al., 2014; Leydesdorff et al., 2014) and subsequently propose cosine similarity as the knowledge proximity. Yan and Luo (2017b) systematically review and comparatively assess various domain-domain proximity measures.

Despite using common structural forms (e.g., citation network), the above-reviewed approaches have specific means for vector representation and proximity calculation. The proximity measures lack interoperability across different entity types and are thus unsuitable for associating heterogeneous pairs, e.g., patent-inventor. Therefore, we adopt a knowledge graph approach, integrating all types of entities and relations from United States Patent and Trademark Office (USPTO) into a single structural form and subsequently embed the knowledge graph onto a unified vector space using various models as reviewed in the following section.

## 2.2. Knowledge Graph Embedding Techniques

Knowledge graph embedding or Knowledge Representation Learning (KRL) is a family of techniques that learn low-rank vector representations of entities and relations that capture the structure and semantics of a knowledge graph (Ji et al., 2021). These techniques vary according to score functions that estimate the plausibility of a fact <h, r, t> relative to other facts in the knowledge graph. Depending on the score functions, the current embedding techniques could be categorized as 'translational distance' and 'semantic matching' models (Zhang et al., 2020). Notable translational distance models include TransE (Bordes et al., 2013), TransR (Lin et al., 2015), and RotateE (Sun et al., 2019). Semantic matching models exploit similarity-based scoring functions, mainly including RESCAL (Nickel et al., 2011), DistMult (Yang et al., 2014), and ComplEx (Trouillon et al., 2016).

Scholars have applied the above-stated models to domain knowledge graphs (e.g., clinical[i]) and subsequently utilized the knowledge graph embeddings for domain tasks. For instance, Mohamed et al. (2021) utilize the embeddings of biological knowledge graphs to predict drug-target interactions and polypharmacy side effects. Huang et al. (2019) train TransE, TransR, and TransH models using the Freebase knowledge graphs – FB2M, FB5M. They apply the learned embeddings to several Question Answering over Knowledge Graph (QA-KG) algorithms that are trained on the SimpleQuestions dataset. Abu-Salih et al. (2021) adopt embedding methods to extract knowledge from social media.

In addition to translational distance and semantic matching models, a variety of models are built using deep learning, specifically using Graph Neural Networks (GNN). Such models, e.g., Graph Convolutional Network (GCN) and its variants, are applied to graphs for supporting domain tasks like semantic relatedness measurement (Mao & Fung, 2020). Such models, however, learn embeddings of an entity using neighborhood and are largely applicable to homogenous graphs. Our work, however, requires graph embedding models for a heterogeneous knowledge graph constructed on patent metadata.

# 3. Method

As shown in Figure 1, we construct a knowledge graph from the US patent database by obtaining the link data that constitutes five types of relations. We train several embedding models using the knowledge graph thus constructed. We then obtain the embeddings of entities and relations that could be associated using cosine similarity.

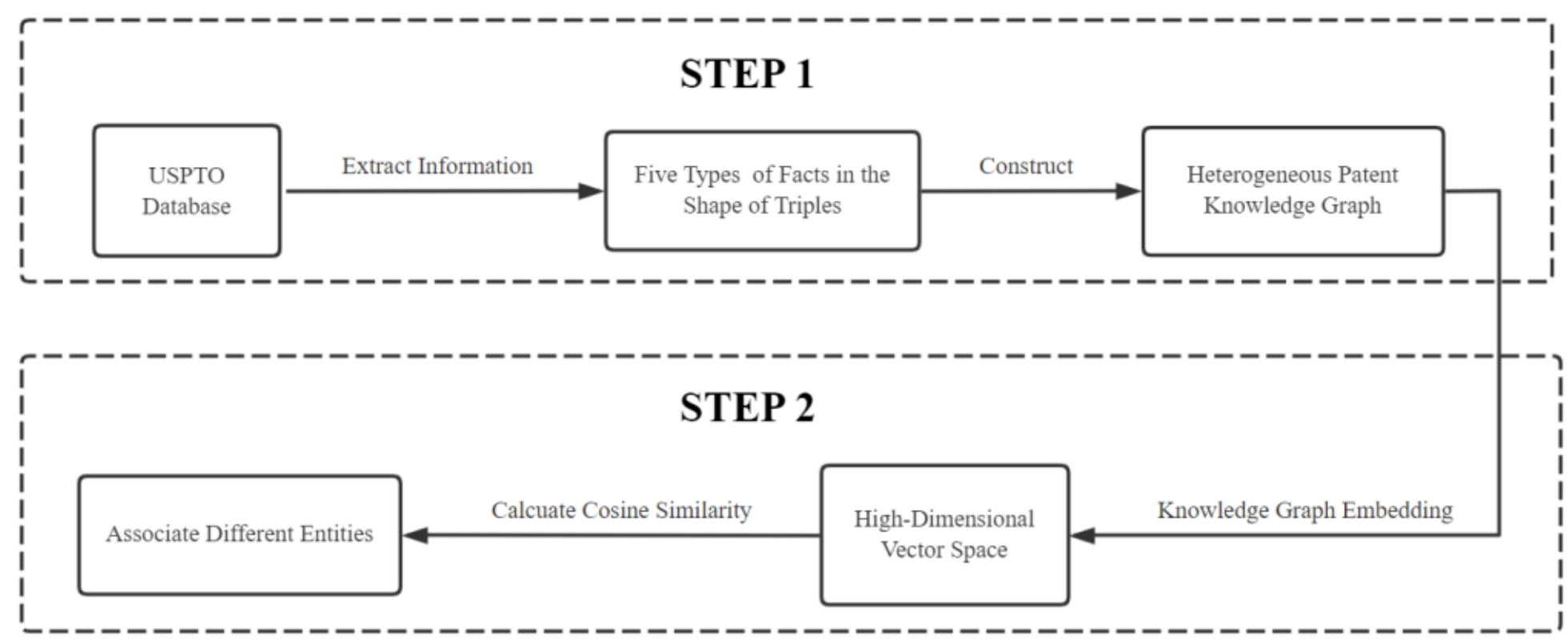

Figure 1. Overview of the proposed method.

## 3.1. Knowledge Graph Construction

We utilize patent metadata from USPTO[ii] to construct a knowledge graph named 'PatNet' that comprises a set of facts {<head entity, relation, tail entity>} where entities belong to 'patent', 'inventor', 'assignee', 'group', and 'subsection' and relations include 'cite', 'write', 'own', 'contain', and 'comprise'. These entities and relations form the following types of facts: <patent, *cite*, patent>, <inventor, *write*, patent>, <assignee, *own*, patent>, <group, *contain*, patent>, and <subsection, *comprise*, groups>. As illustrated in Figure 2, the patent – 'Reverse polysilicon CMOS fabrication' (Patent Number – 5252504) is directly linked to the inventor – Tyler A. Lowrey, and the assignee – Micron Technology Inc., while being classified in the domain 'H01L' (Semiconductor Devices), which is a sub-domain of 'H01' (Basic Electric Elements). In addition, the patent has made and received multiple citations, a couple of which is indicated in Figure 2.

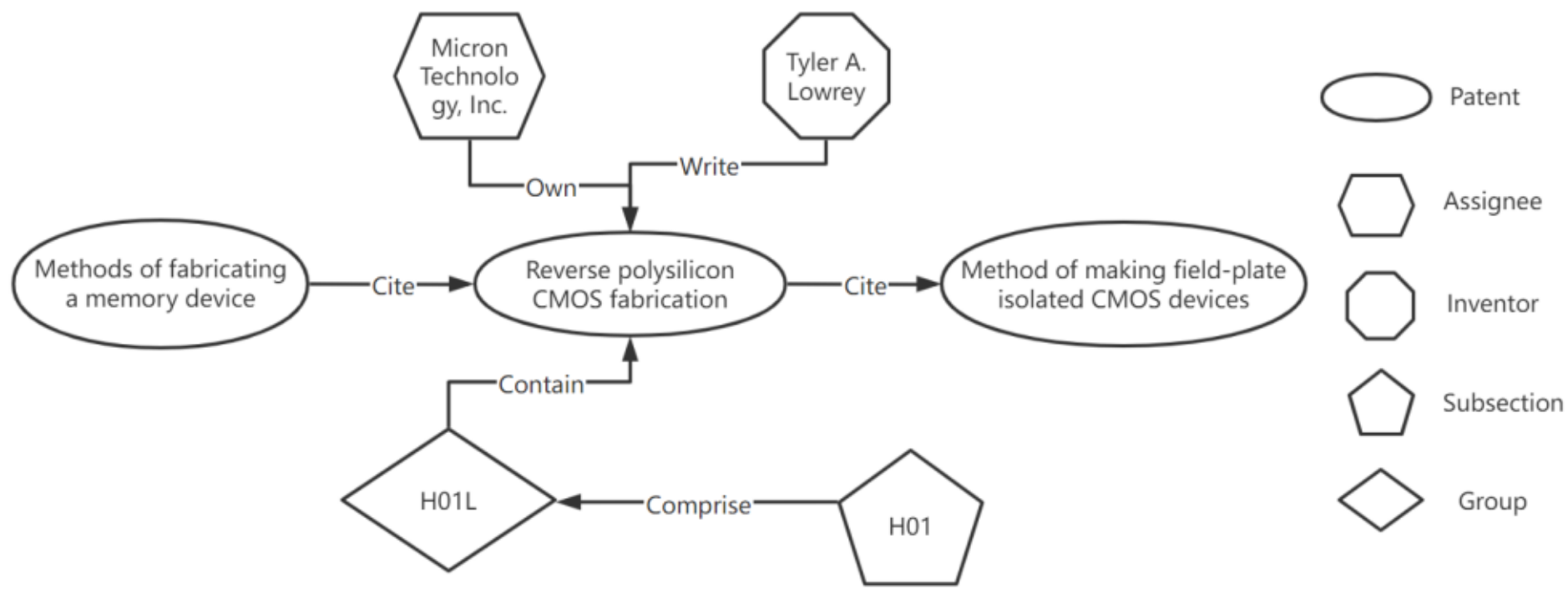

Figure 2: Illustrating entities and relations in PatNet.

The above relations in PatNet amount to 10,273,843 entities and 106,882,276 links. Among the entities, there exist 129 unique subsections given by 3-digit codes (e.g., F02), 667 unique groups given by 4-digit codes (e.g., A03A), 419872 unique assignees, 6037493 unique patents, and 3815682 unique inventors[iii]. Among the relations, there exist 72,724,665 citation links, 6,236,860 own links, 15,852,086 write links, 12,067,998 contain links, and 667 comprise links. Owing to the nature of facts, i.e., unidirectional, PatNet does not include cycles or two-way relations.

## 3.2. Knowledge Graph Embedding

We train the following models using PatNet: TransE_l1, TransE_l2, TransR, RESCAL, DistMult, ComplEx, and RotateE. We have summarized these in Table 1, wherein, h → head entity embedding, t → tail entity embedding, r → relation embedding, d → dimensionality of embedding, and $M_r$ → relationship matrix, R → real vector space, C → complex vector space, Re() → getting the real part of the complex number, diag() → getting diagonal elements on a matrix, $\circ$ → Hadamard product. We explain the notations, score functions, and other limitations of these models in APPENDIX I.

Table 1: Summary of knowledge graph embedding models.

| Method | Entity Embedding | Relation Embedding | Score Function | Complexity |
|---|---|---|---|---|
| TransE | $h, t \in R^d$ | $r \in R^d$ | $-\| h + r - t \|$ | O(d) |
| TransR | $h, t \in R^d$ | $r \in R^k, M_r \in R^{k\times d}$ | $-\| M_r h + r - M_r t \|_2^2$ | O($d^2$) |
| RESCAL | $h, t \in R^d$ | $M_r \in R^{d\times d}$ | $h^T M_r t$ | O($d^2$) |
| DistMult | $h, t \in R^d$ | $r \in R^d$ | $h^T \mathrm{diag}(r) t$ | O(d) |
| ComplEx | $h, t \in C^d$ | $r \in C^d$ | $\mathrm{Re}(h^T \mathrm{diag}(r) t)$ | O(d) |
| RotateE | $h, t \in C^d$ | $r \in C^d$ | $\| h \circ r - t \|$ | O(d) |

To train the models mentioned above, we use DGL-KE[iv] – Deep Graph Knowledge Embedding Library and a server with the following configuration: 8 x NVIDIA Tesla P100-SXM2-16G GPUs and 512 Gigabytes of Memory. To accommodate hardware and time constraints, we set the embedding dimension to 500, preferably between 50 and 1000 (Hogan et al., 2021). While training, the package automatically creates negative facts ($\{\langle h', r, t\rangle\}$ or $\{\langle h, r, t'\rangle\}$) for each positive fact ($\langle h, r, t\rangle$). The package trains a model such that the scoring function is maximized for positive facts and minimized for negative facts. Once the models are trained, we estimate their performances using the test dataset – $\mathcal{D}_{test}$ as follows.

# 4. Evaluation

## 4.1. Predicting Target Entities

We compare the above-trained models using the task of predicting target entities ($\langle ?, r, t\rangle$ or $\langle h, r, ?\rangle$) to assess whether the models capture the structure and semantics of PatNet. In the context of our work, we examine whether the models can predict patent, inventor, domain, or assignee in missing triples, e.g., group in which the new patent could be classified; i.e., identifying the missing entity in the triple <??, *contain*, new patent>. For this evaluation, we arbitrarily gather 10% of the triples in the graph $\mathcal{D}$ as $\mathcal{D}_{test}$, where $|\mathcal{D}_{test}| = 10{,}688{,}227$. For each true triple – $\langle h, r, t\rangle$ in $\mathcal{D}_{test}$ we generate 10,000 possible corrupt triples $\{\langle h', r, t\rangle\}$ and $\{\langle h, r, t'\rangle\}$ by disrupting the head and tail entities. For each true triple – $\langle h, r, t\rangle$ in $\mathcal{D}_{test}$, we identify the rank (in the interval [1, 10,001]) of target entity $h$ or $t$ amidst the corrupt entities $h'$ or $t'$ that are present in corrupt triples $\{\langle h', r, t\rangle\}$ and $\{\langle h, r, t'\rangle\}$.

Based on these ranks, we compute the following metrics that are elaborated in APPENDIX II.

1. Mean Rank (MR) is the average of the ranks of all target entities in the $\mathcal{D}_{test}$. This metric lies in the interval $[1, 10{,}001]$ and is often found to be highly sensitive to outliers whose rank is $\gg 1$.
2. Mean Reciprocal Rank (MRR) is the average inverse of the ranks of all target entities in the $\mathcal{D}_{test}$. This metric is less sensitive to outliers and lies in the interval $(0,1]$.
3. Hits@k represents the proportion of target entities in $\mathcal{D}_{test}$ whose rank $\leq k$.

The higher performance of an embedding model in a link prediction task is given by higher MRR, higher hits@k, and lower MR. We report the above-stated metrics for all embedding models in Table 2.

Table 2: Summary of the performances of embedding models in predicting target entities.

| | MR | MRR ↓ | HITS@1 | HITS@3 | HITS@10 |
|---|---|---|---|---|---|
| RESCAL | 6.210 | 0.928 | 0.905 | 0.947 | 0.958 |
| ComplEx | 6.254 | 0.911 | 0.879 | 0.938 | 0.955 |
| DistMult | 6.204 | 0.906 | 0.870 | 0.937 | 0.955 |
| TransE_l2 | 6.634 | 0.888 | 0.842 | 0.927 | 0.951 |
| RotateE | 39.027 | 0.762 | 0.687 | 0.819 | 0.886 |
| TransR | 55.835 | 0.654 | 0.579 | 0.698 | 0.787 |
| TransE_l1 | 252.603 | 0.626 | 0.545 | 0.680 | 0.770 |

While RESCAL returns the highest MRR and hits@k, we can distinguish the performances into two categories. The first category comprising RESCAL, ComplEx, DistMult, and TransE_l2 exhibits not only closer MR [6.204, 6.254], but also lies within a closer range in MRR – [0.888, 0.928]. The gap amongst these models is further narrowed in Hits@10 – [0.951, 0.958]. Across all metrics, the second category of models comprising RotateE, TransR, and TransE_l1 are quite distinguishable from the first category, which better captures the structure and semantics of PatNet. Given a source entity $h$ or $t$, the models in the first category can better predict the associated entity $t$ or $h$ through a specific relation $r$ that could be *write*, *comprise*, *own*, *cite*, or *contain*.

## 4.2. Assessing Domain Expansion Profiles

Literature suggests that assignees and inventors often diversify their portfolios by exploring technological domains that are less distant from their prior domains (Alstott et al., 2017a). In this section, we study the domain expansion history of all inventors and assignees (commonly referred to as 'agents' henceforth) in USPTO to examine how well these are explainable by the proposed knowledge proximity measure. As shown in the synthetic example in Figure 3, let us consider an agent who holds patents in three domains – A, B, and C (together constituting the home domain) and could explore target domains – D, E, and F.

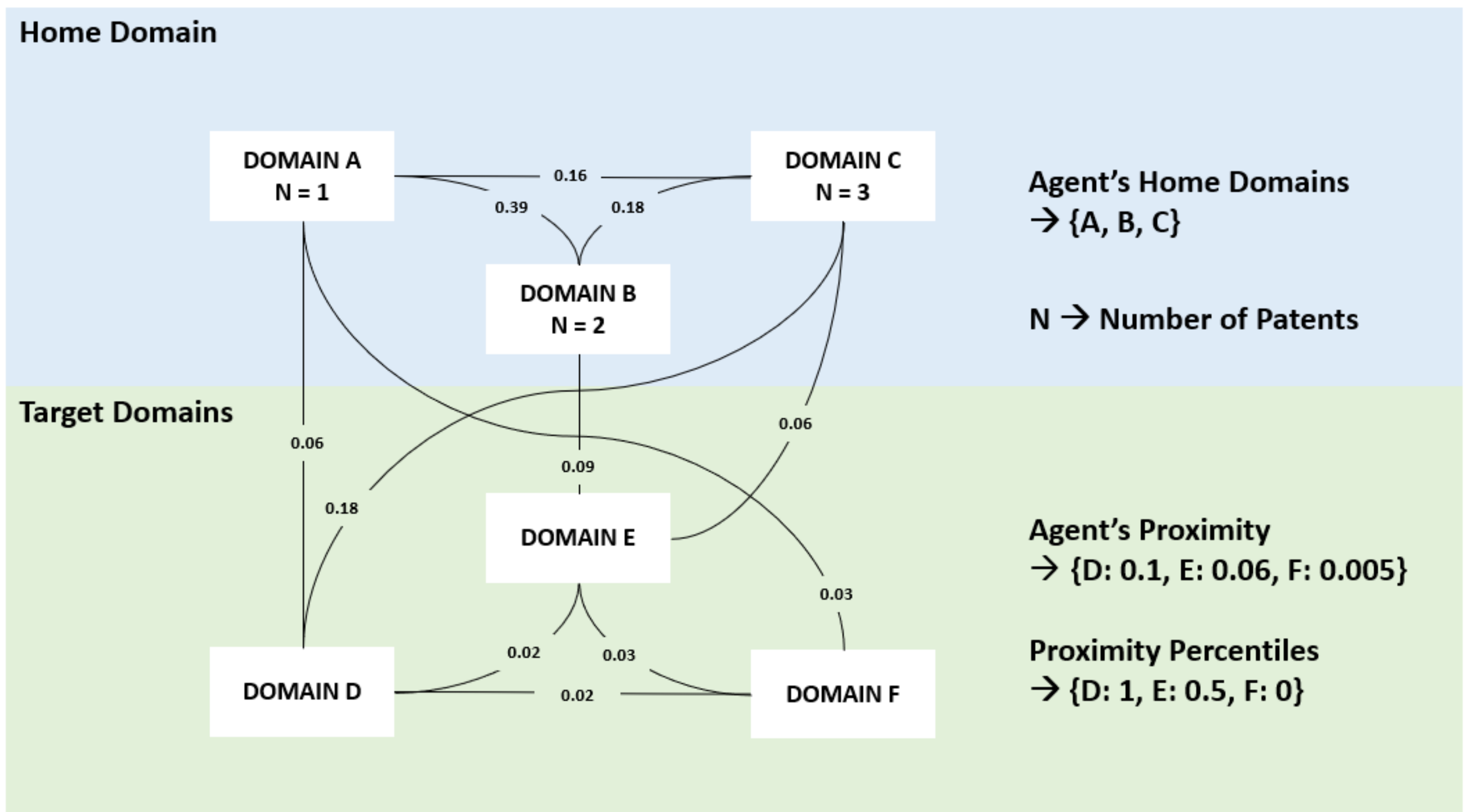

Figure 3: Illustrating calculation of overall proximity between the home domain and a target domain.

The likelihood of the agent entering a target domain, given by Eq. 1, is given by the average of individual proximities to all home domains, weighted by the number of patents in these.

$$Domain - Agent\ Proximity(a, j) = \frac{\sum_{i \neq j} \varphi_{ij}\, a_i}{\sum_{i \neq j} a_i} \tag{1}$$

Where $\varphi_{ij}$ denotes the knowledge proximity (cosine similarity between the domain embeddings) between domains $i$ and $j$, that belong to, respectively, home and target domains; $a_i$ represents the number of patents belonging to the field $i$ owned by the agent $a$. The above equation yields the values 0.1, 0.06, and 0.005 as proximities to the domains D, E, and F respectively. We obtain proximity percentiles according to these scores, i.e., 1, 0.5, and 0 to identify the closest domain. As the agent enters new domains, we mark the entry to a domain as its proximity percentile with respect to the instantaneous home domain. In this example, if the agent enters target domains in the order D-E-F, the expansion profile could be represented as (1, 1, 1), while indicating that the proximity measure explains the whole expansion profile.

In a general form, the expansion profile of an agent could be written as ($pp_1, pp_2, pp_3$ ...), where $pp_i$ stands for the proximity percentile of the $i^{th}$ domain during the time of entry. As illustrated in Figure 4, the profiles of multiple agents could be concatenated into a combined expansion profile. While these profiles differ according to the embedding model, the preferred choice of model should maximize the proximity percentiles in the combined profile. We plot the cumulative distribution of proximity percentile for three pseudo models in Figure 4.

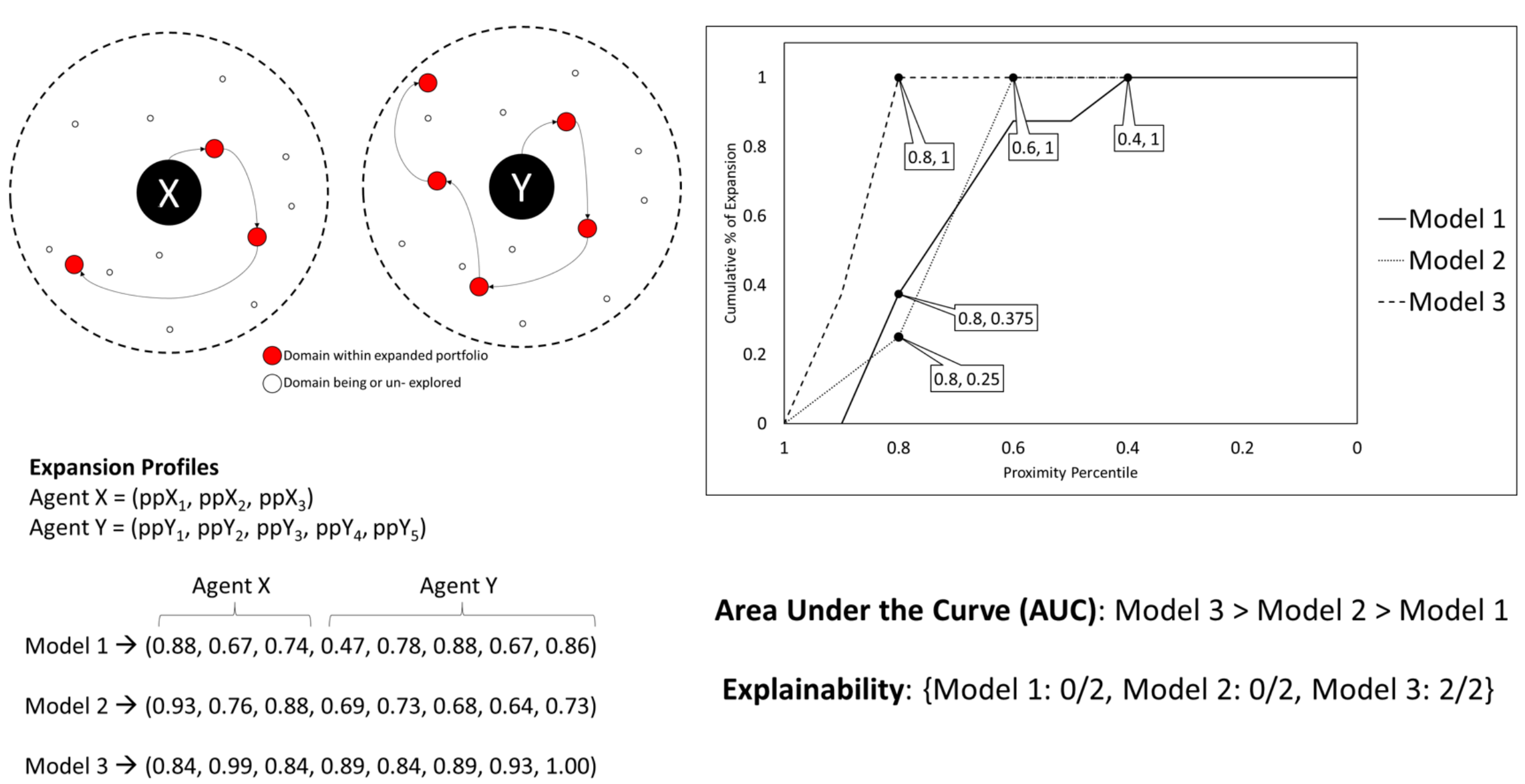

Figure 4: Illustrating the comparison of different models in terms of AUC and explainability.

The distribution represents the proportion of the expansion profile that lies above a proximity percentile. For instance, the data point (0.8, 1) for Model 3 indicates that the entire expansion profile has a proximity percentile above 0.8, which could be verified alongside. This implies that at every step of domain expansion, the agent has entered a domain that was ranked among the first 20% by Model 3. Such a distribution should enclose a higher Area Under the Curve (AUC) compared to Models 1 and 2. A preferred model should exhibit higher AUC with respect to individual profiles (e.g., X and Y) as well. In that regard, Model 3 exhibits higher AUC for both profiles, indicating an 'explainability' of 2/2 = 1.

We measure AUC and explainability as the performance metrics for the combined expansion profiles of 76,326 inventors and 15,283 assignees with at least 30 patents in the US patent database. For every agent, we serialize the patents according to the application date. For every patent in the sequence, we form an instantaneous set of home and target domains[v] (refer to Figure 3 for example) and rank the target domains using proximity percentile. Using the next patent in the sequence, we identify the proximity percentiles of entered domains and append these to the expansion profile (refer to Figure 4 for example). Thus, we obtain the individual expansion profiles of all agents, which we then concatenate to form combined profiles for inventors and assignees.

Upon computing the expansion profiles using the seven embedding models, as shown in Figure 5, we plot the cumulative distribution of proximity percentiles. TransE_l2 exhibits the highest AUC for the combined expansion profiles of both assignees and inventors, indicating better predictability at each step of domain expansion. As shown in Figure 6, TransE_l2 also returns the highest AUC for nearly 70% of assignees and inventors, which is significantly better than TransE_l1 ($\approx 20\%$) and other models ($< 10\%$). TransE_l2, therefore, shows higher explainability of combined profiles of assignees and inventors.

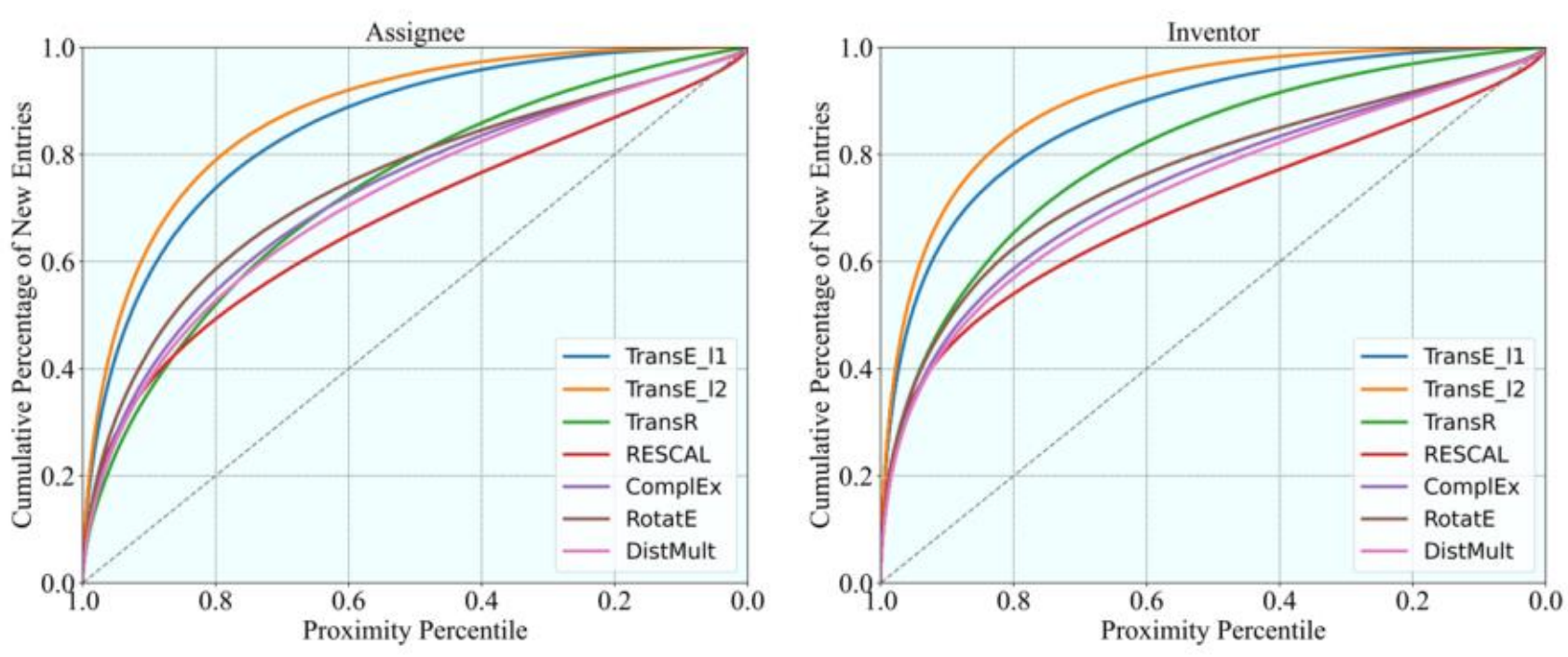

| | Assignee AUC | Inventor AUC |
|---|---|---|
| **TransE_l2** | 0.875 | 0.902 |
| **TransE_l1** | 0.851 | 0.871 |
| **RotateE** | 0.743 | 0.760 |
| **TransR** | 0.729 | 0.803 |
| **ComplEx** | 0.725 | 0.740 |
| **DistMult** | 0.714 | 0.730 |
| **RESCAL** | 0.675 | 0.696 |

Figure 5: Cumulative distributions (including AUC) of the proximity percentiles.

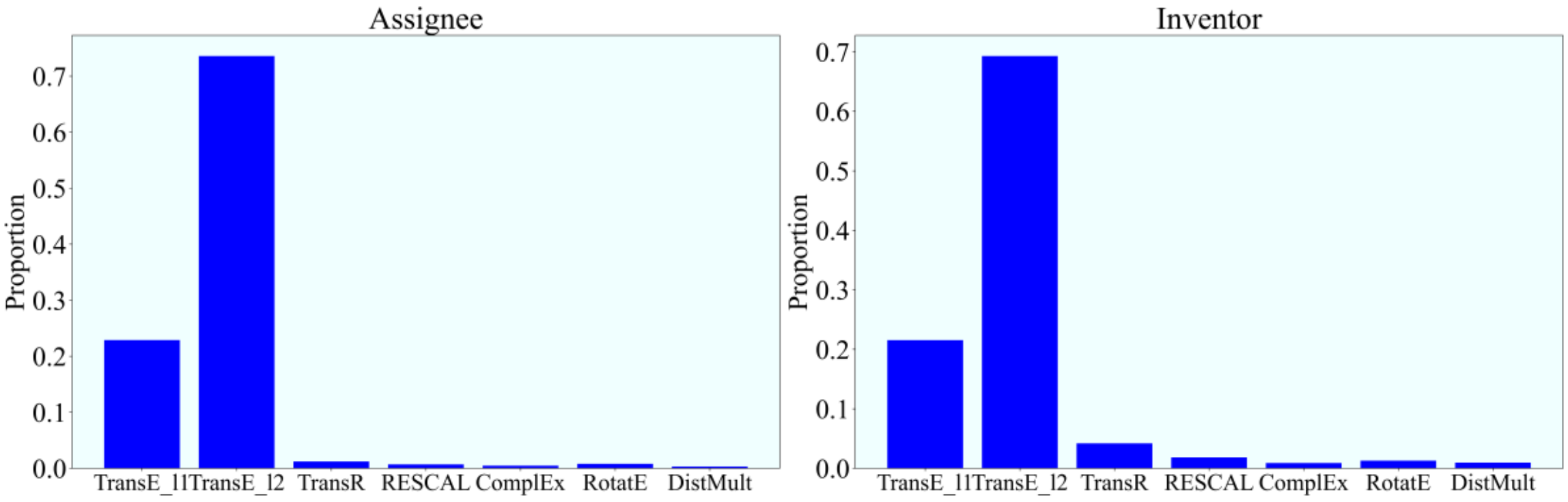

Figure 6: Explainability of each model that is measured as the proportion of the number of agents (inventor or assignee) where the model exhibits higher AUC compared to other models.

Based on the well-understood premise that inventors and assignees expand their portfolio by entering less distant domains, we compared the embedding models in terms of explaining their expansion profiles, as given by USPTO. TransE_l2 embeddings offer better performance in this task as well as in the task of predicting target entities (Section 4.1) alongside RESCAL, ComplEx, and DistMult. As the latter models exhibit poor

AUC and explainability in this task, we could infer that TransE_l2 better captures the structure and semantics of PatNet while also forming meaningful associations among entities. While TransE_l2 embeddings shall be utilized for PatNet-related applications, it is important to note that other models could exhibit superior performance in specific domain task(s) that could be envisaged by scholars.

## 5. Applications of Knowledge Proximity

In this section, we demonstrate how PatNet embeddings could be utilized to associate homogenous and heterogeneous pairs of entities. For associating homogeneous pairs of entities (e.g., patent-patent), it is possible to directly compute the cosine similarity between the embeddings of these. For heterogeneous pairs (e.g., patent-inventor), however, it is necessary to transform target entities into the type of focal entity as follows.

$$h + r \approx t \tag{2}$$

The above transformation denotes that the sum of embeddings of a head entity $h$ and the relation $r$ approximates a tail entity $t$. The sum of the embeddings of assignee-X and the relation – 'own', for instance, could result in a patent-X, which is not an actual patent but a patent equivalent of assignee-X. Once the entities are transformed to a single type, it is possible to calculate cosine similarity between the transformed embeddings. Based on the types of triples in PatNet, in Table 3, we provide the guide to transforming entities into other types.

Table 3: Guide to transforming target entities. The operator emb(.) refers to the embedding of the operand.

| Focal Entity / Target Entity | Patent | Inventor | Assignee | Domain – Group | Domain – Subsection |
|---|---|---|---|---|---|
| Patent | No Transformation | emb(target) – emb('write') | emb(target) – emb('own') | emb(target) – emb('contain') | emb(target) – emb('contain') – emb('comprise') |
| Inventor | emb(target) + emb('write') | No Transformation | emb(target) + emb('write') – emb('own') | emb(target) + emb('write') – emb('contain') | emb(target) + emb('write') – emb('contain') – emb('comprise') |
| Assignee | emb(target) + emb('own') | emb(target) + emb('own') – emb('write') | No Transformation | emb(target) + emb('own') – emb('contain') | emb(target) + emb('own') – emb('contain') – emb('comprise') |
| Group | emb(target) + emb('contain') | emb(target) + emb('contain') – emb('write') | emb(target) + emb('contain') – emb('own') | No Transformation | emb(target) – emb('comprise') |
| Subsection | emb(target) + emb('comprise') + emb('contain') | emb(target) + emb('comprise') + emb('contain') – emb('write') | emb(target) + emb('comprise') + emb('contain')– emb('own') | emb(target) + emb('comprise') | No Transformation |

According to Table 3,

1. Transformation is not required for the target entities of the same type as the focal entity, e.g., assignee-assignee.
2. The transformation could be accomplished in a single step when focal and target entities are directly connected by one of the relations ('write', 'own', 'contain', and 'comprise'), e.g., inventor-patent.
3. The transformation could require multiple steps when the focal and target are indirectly connected through more than one relation, e.g., assignee-inventor.

Guided by Table 3, we apply the embeddings of TransE_l2 for 1) visualizing patent embeddings, 2) exploring the nearest neighborhood of an entity, and 3) examining a system of heterogeneous entities. Upon demonstrating these applications, we also speculate on various use cases for future academic and business tools. (**Note**: The TransE_l2 embeddings of all entities and relations, along with the associated data, are available in OneDrive[vi]).

### 5.1. Visualizing Patent Associations (Homogeneous Pairing)

To demonstrate how well the patent embeddings capture the classification system of USPTO, we cluster the embeddings of patents from various domains. Since this application involves only patents, transformation is not necessary. For visual clarity, we obtain patents from 20 domains (e.g., D21D) whose patent numbers range from 1,200 to 1,400. We apply the t-distributed stochastic neighbour embedding (T-SNE) method to reduce the dimensionality of patent embeddings and visualize these in a 2-D plot (Figure 7), wherein, the clusters primarily represent the classification scheme of the patent database. Occasionally, the patents that belong to the same domains have been distributed among different clusters. Such cases could potentially indicate that pairwise knowledge proximities among patents capture other associations like citations and ownership in PatNet.

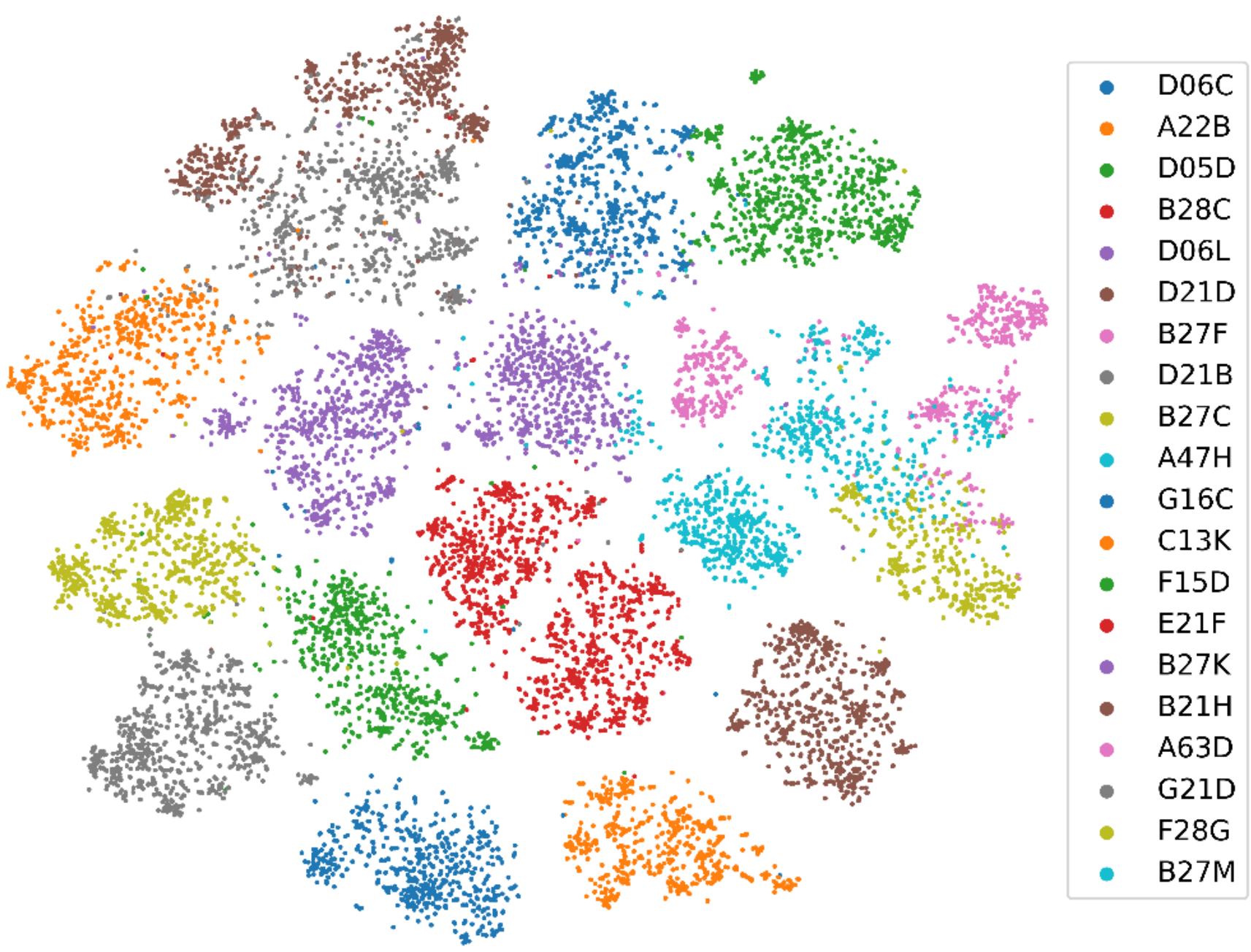

Figure 7: Visualizing patent embeddings across 20 domains.

### 5.2. Exploring Nearest Neighborhood (Heterogeneous Pairing)

In this application, we retrieve entities closest to a focal entity and form the nearest neighborhood. Since this application involves all kinds of entities, it is necessary to transform all target entities in USPTO into the type of focal entity. In Figure 8, for a focal entity – a patent titled "Entropy coding scheme for video coding" (Patent Number – 7158684), we visualize a few directly associated entities as well as indicate the knowledge proximity (link weights) values. Despite having no direct association, the inferred link (dashed line) weight of 0.717 between "Yuji Itoh" and "Ngai-Man Cheung" suggests that the embeddings capture the extended association between these inventors. These directly associated entities are not the closest to the focal entity, e.g., knowledge proximity = 0.389 to Texas Instrument Incorporated.

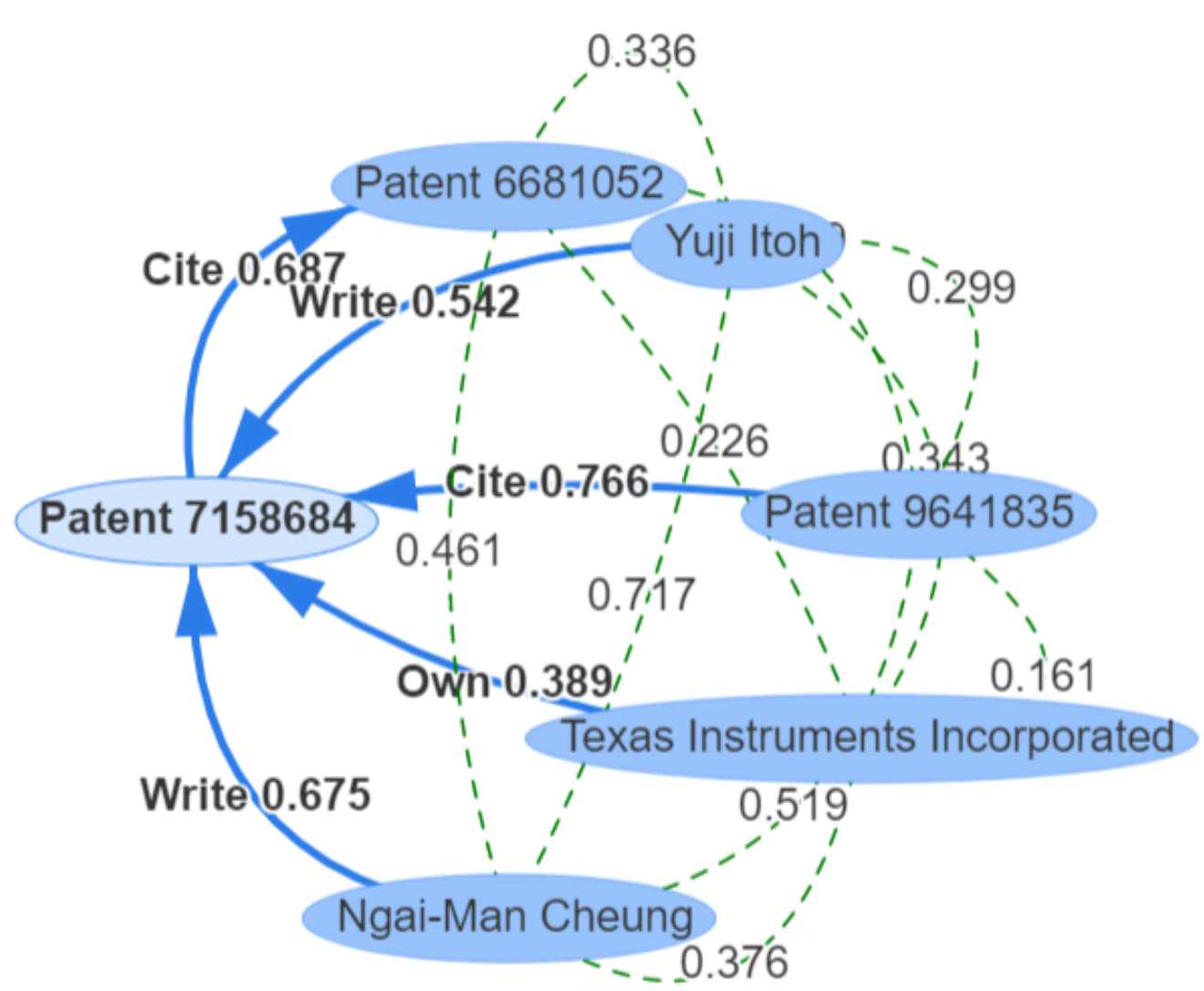

Figure 8: Knowledge proximity in heterogeneous pairs

Upon transforming all target entities in USPTO to inventor type, we identify those that are closest to the inventor – 'Dawn Tan' in the total PatNet embedding space. As shown in Table 4a, the closest target entity is the patent – 9971091, for which the focal entity – Dawn Tan, is the sole inventor. Another patent – 9671673 of Dawn Tan, including the other inventors, is part of the neighborhood. Both patents in the neighborhood are assigned to the Singapore University of Technology and Design, which is also among the five closest entities to Dawn Tan.

Table 4a: The top five entities closest to the focal entity – an inventor named "Dawn Tan".

| Rank | Target Entity | Target Entity Type | Knowledge Proximity to the focal entity | Relation with the focal entity |
|---|---|---|---|---|
| 1 | Optical devices and methods for fabricating an optical device (Patent Number - 9971091) | Patent | 0.908 | Dawn Tan is the sole inventor of Patent 9971091 |
| 2 | Christine Donnelly | Inventor | 0.896 | Dawn Tan and Christine Donnelly are co-inventors of patent – 9671673 |
| 3 | George F. R. Chen | Inventor | 0.885 | Dawn Tan and George F. R. Chen are co-inventors of patent – 9671673 |
| 4 | Optical device for dispersion compensation (Patent Number - 9671673) | Patent | 0.878 | Dawn Tan is one of the inventors of patent – 9671673 |
| 5 | Singapore University of Technology and Design | Assignee | 0.718 | Dawn Tan's patents 9971091 and 9671673 are both assigned to Singapore University of Technology and Design |

In a similar approach, we also explore the nearest neighborhood of the patent – 9971091 (Table 4b). Among the five closest entities, the inventor – Dawn Tan is ranked first with a proximity = 0.892 to the focal entity. The knowledge proximity between the inventor – Dawn Tan, and the patent – 9971091, is indicated as 0.908 and 0.892 in Tables 4a and 4b, respectively. The variation in proximity score is due to the difference in transformation adopted, i.e., the difference in the type of focal entity (inventor and patent) in these examples.

Table 4b: The top five entities closest to Patent 9971091.

| Rank | Target Entity | Target Entity Type | Knowledge Proximity to the focal entity (Patent – 9971091) | Relation with the focal entity (Patent – 9971091) |
|---|---|---|---|---|
| 1 | Dawn Tan | Inventor | 0.892 | Focal patent's inventor |
| 2 | Optical device for dispersion compensation (Patent Number - 9671673) | Patent | 0.841 | This patent shares the same inventor and assignee with the focal patent and is also cited by the focal patent |
| 3 | Christine Donnelly | Inventor | 0.801 | This inventor is also the co-inventor with the focal patent's inventor but for a different patent |
| 4 | George F. R. Chen | Inventor | 0.778 | This inventor is also the co-inventor with the focal patent's inventor but for a different patent |
| 5 | Singapore University of Technology and Design | Assignee | 0.754 | SUTD is the assignee of the focal patent |

In Table 4b, the patent – 9671673 is ranked as the second closest to the focal patent. While patent – 9971091 cites patent – 9671673, these two patents also share a joint inventor (Dawn Tan), an assignee (Singapore University of Technology and Design), and a classification (G02B – Optical Elements, Systems or Apparatus). Among these associations, the assignee – Singapore University of Technology and Design, is ranked among the five closest entities to the patent – 9971091. A joint interpretation from the two neighborhoods (Table 4a & 4b) is that the retrieved entities form a tightly coupled network, involving technical collaborations and organizational ties.

## 5.3. Examining Pairwise Proximities (Heterogeneous Pairing)

In this section, we examine a closed system of entities comprising patents, domains, assignees, and inventors associated with knowledge proximity. As illustrated in Figure 9, we consider a system that includes H04L (group), Massachusetts Institute of Technology (assignee), Singapore University of Technology Design (assignee), Microsoft Corporation (assignee), Hui Ying Yang (inventor), Kristin Wood (inventor), Pablo A.

Valdivia Y Alvarado (inventor), “Touch Screen Video Gaming Machine” (patent), and “Global Hosting System” (patent).

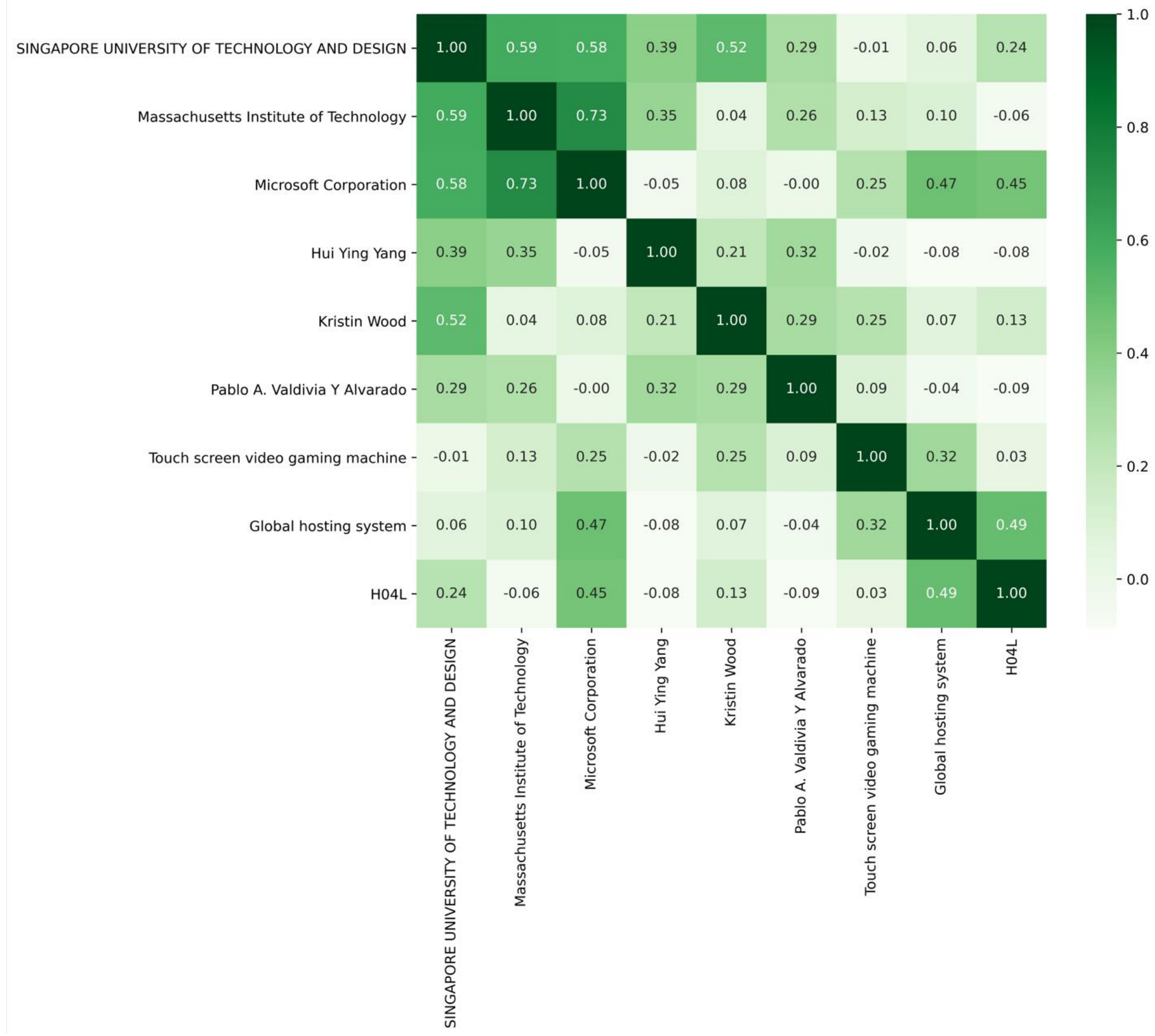

Figure 9: Pairwise knowledge proximities among a heterogeneous set of entities.

To compute the knowledge proximity values (as indicated in Figure 9), we transform all entities (based on Table 3) to patent equivalents. Based on the knowledge proximity values, we understand that all three inventors have higher proximities with “Singapore University of Technology and Design” (SUTD), which is closer to “Massachusetts Institute of Technology” (MIT) compared to Microsoft Corporation. This suggests that the assignees SUTD and MIT could potentially hold organizational ties. In addition, these presumed ties could be mediated by two of the three inventors who appear closer to both assignees.

While Xbox is one of the projects of Microsoft Corporation, it is not uncommon to assume that the patent “Touch Screen Video Gaming Machine” could be closer to Microsoft Corporation than the patent “Global Hosting System.” Contrary to popular opinion, Figure 9 informs that Microsoft Corporation specializes in network architectures more than gaming consoles. This observation is also consistent with the proximity between Microsoft Corporation and the group – H04L that denotes the transmission of digital information. These inferences could help perceive organizations, their members, and possible collaborations.

# 6. Conclusions and Future Work

In this article, we have operationalized knowledge proximity within the context of the US Patent database that comprises various entities such as patents, inventors, assignees, and domain classifications. We integrated patent metadata from USPTO into a single knowledge graph called PatNet that comprises 106,882,276 triples that constitute five types of relations: <patent, *cite*, patent>, <inventor, *write*, patent>, <assignee, *own*, patent>, <group, *contain*, patent> and <subsection, *comprise*, groups>. We trained seven graph embedding models using PatNet and identified that RESCAL, DistMult, ComplEx, and TransE_l2 exhibit satisfactory performance in terms of predicting target entities ($\langle ?, r, t \rangle$ or $\langle h, r, ? \rangle$). In terms of explaining the expansion profiles of all assignees and inventors, TransE_l2 exhibits higher AUC and explainability. Based on the results, we applied TransE_l2 embeddings to cluster patents, retrieve the nearest neighborhood of a focal entity and examine a system of heterogeneous entities.

Our method is limited in the following aspects that indicate potential research opportunities for the future development of our work.

- Knowledge graphs consume higher memory compared to the link tables in the patent database while lacking the sufficient infrastructure to reduce run time.
- Compared to traditional representations like regular networks, the knowledge graph representation is relatively challenging to comprehend and explain (using network properties).
- The current knowledge proximity may be less applicable to highly focused studies, as proximity is based on all entities and relations in USPTO. Learning the embeddings by conducting training on the desired subset of PatNet could alleviate this issue.
- The facts in PatNet are captured without relative importance, e.g., in the facts <inventor1, *write*, patent1> and <inventor2, *write*, patent1>, the relation *write* carries a similar level of importance irrespective of the inventor contributions.
- Since PatNet is restricted to USPTO, it would be extended to other patent databases, research articles, and other technical publications.

The approach to measuring knowledge proximity, based on PatNet and the knowledge graph embeddings, enables homogenous and heterogeneous associations among inventions, people, organizations, and technological fields, as demonstrated using various examples in this article. Such associations also help perceive these entities and make inferences thereupon. Knowledge proximity could thus be utilized as a fundamental instrument for the development of various patent-related applications and eventually contribute to data-driven innovation, business, and policy intelligence (Luo, 2022; Sarica et al., 2020).

# APPENDIX I

*TransE and Its Extensions*

Translation-based embedding model (TransE) is a representative translational distance model that represents entities and relations as vectors in the same vector space of dimension $\mathbb{R}_d$, where $d$ is the dimension of the target space with reduced dimension. TransE performs the linear transformation of vectors by adding a relation $r$ to a head $h$ to approximate the tail $t$ in a knowledge graph triplet fact.

$$h + r \approx t \tag{1}$$

For example, if $h_{Patent_1} = emb('Patent_1')$, $r_{belongs_to} = emb('belongs_to')$, $t_{H04L} = emb('H04L')$, then $h_{Patent_1} + r_{belongs_to}$ should approximate $t_{H04L}$. The scoring function of TransE is negative distance between $h + r$ and $t$:

$$f = -\|h + r - t\|_{\frac{1}{2}} \tag{2}$$

TransE_l1 and TransE_l2 are two extensions of TransE. TransE_l1 uses L1 regularization that is calculated as the sum of the absolute values of the vector, while TransE_l2 uses L2 regularization that is calculated as the square root of the sum of the squared vector values.

*TransR*

Both TransE and TransR are called translational distance models as they translate the entities, relations and measure distance in the target vector spaces. Different from TransE that embeds entities and relations into a same dimensional vector space, TransR separates entity space from relation space where $h, t \in \mathcal{R}^k$ and $r \in \mathcal{R}^d$. A projection matrix $M \in \mathcal{R}^{k*d}$ is learned for each relation to project the entities to the relation space. The projection matrix $M_r$ allows TransR to deal with the relation that is not 1-to-1 as each $M_r$ captures the features of a certain relation. Similar with TransE, TransR uses the same scoring function of measuring the Euclidean distance between $h + r$ and $t$ in certain relation space, $h_r = hM_r$ and $t_r = tM_r$, more normally, $f_r = -\| h_r + r - t_r \|_{\frac{1}{2}}$.

*RESCAL, DistMult and ComplEx*

RESCAL is a bilinear model that uses vectors to represent entities, matrices to represent relations, and a custom scoring function to capture the internal interactions of triples. RESCAL captures the structure information of the knowledge graph by using a three-dimensional tensor $\mathcal{X}$ that models pairwise interaction between entities. $\mathcal{X}_{ijk}$ contains the fact between the $i^{th}$ entity and the $j^{th}$ entity under the $k^{th}$ relation. Value of $\mathcal{X}_{ijk}$ is defined as:

$$\mathcal{X}_{ijk} = \begin{cases} 1 & \text{if } (e_i, r_k, e_j) \text{ holds} \\ 0 & \text{if } (e_i, r_k, e_j) \text{ does not hold} \end{cases} \tag{3}$$

For a graph with many entities, $\mathcal{X}$ can be sparse and asymmetrical. RESCAL decomposes each layer of $\mathcal{X}$ to capture the inherent graph structure in the form of a latent vector representation of the entities and an asymmetric square matrix that captures the relationships. The decomposition progress is defined as follow:

$$\mathcal{X}_k \approx AR_kA^{\top}, \text{ for } k = 1, \dots, m \tag{4}$$

where $A$ is an $n \times r$ matrix of latent component representation of entities, and the asymmetrical $R_k$ is an $r \times r$ square matrix that represents the interaction for $k^{th}$ predicate entity in $\mathcal{X}$. $m$ is the number of entities and relations respectively. $A$ and $R_k$ are computed through minimizing the distance between $\mathcal{X}_k$ and $AR_kA^{\top}$. RESCAL uses a similarity-based scoring function that measures the credibility of facts by matching the underlying semantics of entities and the relations contained in the vector space representation. The scoring function is bilinear.

$$f(A, R_k) = \frac{1}{2}\sum_{i,j,k} \left(\mathcal{X}_{ijk} - \mathbf{a}_i^T R_k \mathbf{a}_j\right)^2 \tag{5}$$

where $a_i$ and $a_j$ are the $i^{th}$ and $j^{th}$ rows of $A$ and thus are the latent-component representations of the $i^{th}$ and $j^{th}$ entity. RESCAL is easy to overfit. The complexity will be high as the dimension of the relation matrix increases, making it difficult to apply to large-scale knowledge graphs.

DistMult uses the diagonal matrix to represent the relationship matrix, reducing the number of parameters of the bilinear model to the same as TransE. However, DistMult oversimplifies the RESCAL model. It can only solve the symmetrical relations in the knowledge graph. ComplEx extends DistMult to the complex number space, so it can solve both symmetric and asymmetric relations at the same time.

*RotateE model*

Inspired by Euler decomposition, the RotateE model maps entities and relations to a complex vector space and defines each relation as a rotation from the head to the tail entity. Given a triplet $(h, r, t)$, $t = h \circ r$ , where $h, r, t \in \mathbb{C}^k$ are the embeddings of the head, relation, and tail in the complex vector space separately, and $\circ$ is the Hadamard product. For each set of $(h_i, r_i, t_i)$, the following relationship is expected.

$$t_i = h_i r_i, \text{ where } h_i, r_i, t_i \in \mathbb{C}, \text{ and } |r_i| = 1 \tag{6}$$

The scoring function of RotateE measures the angular distance and is defined as:

$$f_r(h, t) = \| h \circ r - t \| \tag{7}$$

Because RotateE remains linear in time and memory, it can be extended to a large knowledge graph.

# APPENDIX II

Three standard metrics are used to evaluate the embedding quality, including Mean Reciprocal Rank (MRR), Mean Rank (MR), and Hit ratio with cut-off values $n$ = 1, 3, and 10. MRR is the average of the reciprocal ranks of results for a sample of queries $Q$. Its value ranges from 0 to 1 as the best. It is given as:

$$\text{MRR} = \frac{1}{|Q|} \sum_{i=1}^{|Q|} \frac{1}{\text{rank}_i} \tag{1}$$

where $rank_i$ refers to the rank position of the positive triplet for the $i^{th}$ query.

MR measures the average rank of all correct entities with a lower value representing better performance. It is given as:

$$\mathrm{MR} = \frac{1}{|Q|}\sum_{i=1}^{|Q|} \mathrm{rank}_i \tag{2}$$

Hits@k describes the fraction of true entities that appear in the first *k* entities of the sorted rank list. Its value lies in (0,1] where closer to 1 is better. It is given as:

$$\mathrm{hits@k} = \frac{1}{|Q|}\sum_{i=1}^{|Q|} \mathbb{I}\left[\mathrm{rank}_i \leq k\right] \tag{3}$$

# References

Abu-Salih, B., Al-Tawil, M., Aljarah, I., Faris, H., Wongthongtham, P., Chan, K. Y., & Beheshti, A. (2021). Relational learning analysis of social politics using knowledge graph embedding. *Data Mining and Knowledge Discovery*, *35*(4), 1497-1536.

Aharonson, B. S., & Schilling, M. A. (2016). Mapping the technological landscape: Measuring technology distance, technological footprints, and technology evolution. *Research policy*, *45*(1), 81-96.

Ahuja, G. (2000). The duality of collaboration: Inducements and opportunities in the formation of interfirm linkages. *Strategic management journal*, *21*(3), 317-343.

Alstott, J., Triulzi, G., Yan, B., & Luo, J. (2017a). Inventors' explorations across technology domains. *Design Science*, *3*.

Alstott, J., Triulzi, G., Yan, B., & Luo, J. (2017b). Mapping technology space by normalizing patent networks. *Scientometrics*, *110*(1), 443-479.

An, X., Li, J., Xu, S., Chen, L., & Sun, W. (2021). An improved patent similarity measurement based on entities and semantic relations. *Journal of Informetrics*, *15*(2), 101135.

Bordes, A., Usunier, N., Garcia-Duran, A., Weston, J., & Yakhnenko, O. (2013). Translating embeddings for modeling multi-relational data. *Advances in neural information processing systems*, *26*.

Deerwester, S., Dumais, S. T., Furnas, G. W., Landauer, T. K., & Harshman, R. (1990). Indexing by latent semantic analysis. *Journal of the American Society for information Science*, *41*(6), 391-407.

Dibiaggio, L., & Nesta, L. (2005). Patents statistics, knowledge specialisation and the organisation of competencies. *Revue d'économie industrielle*, *110*(1), 103-126.

Diestre, L., & Rajagopalan, N. (2012). Are all 'sharks' dangerous? new biotechnology ventures and partner selection in R&D alliances. *Strategic management journal*, *33*(10), 1115-1134.

Feng, S. (2020). The proximity of ideas: An analysis of patent text using machine learning. *PloS one*, *15*(7), e0234880.

Gerken, J. M., & Moehrle, M. G. (2012). A new instrument for technology monitoring: novelty in patents measured by semantic patent analysis. *Scientometrics*, *91*(3), 645-670.

Guan, J. C., & Yan, Y. (2016). Technological proximity and recombinative innovation in the alternative energy field. *Research policy*, *45*(7), 1460-1473.

Hogan, A., Blomqvist, E., Cochez, M., d'Amato, C., Melo, G. d., Gutierrez, C., . . . Neumaier, S. (2021). Knowledge graphs. *Synthesis Lectures on Data, Semantics, and Knowledge*, *12*(2), 1-257.

Huang, X., Zhang, J., Li, D., & Li, P. (2019). Knowledge graph embedding based question answering. Proceedings of the twelfth ACM international conference on web search and data mining,

Ji, S., Pan, S., Cambria, E., Marttinen, P., & Philip, S. Y. (2021). A survey on knowledge graphs: Representation, acquisition, and applications. *IEEE Transactions on Neural Networks and Learning Systems*, *33*(2), 494-514.

Kay, L., Newman, N., Youtie, J., Porter, A. L., & Rafols, I. (2014). Patent overlay mapping: Visualizing technological distance. *Journal of the Association for Information Science and Technology*, *65*(12), 2432-2443.

Leydesdorff, L., Kushnir, D., & Rafols, I. (2014). Interactive overlay maps for US patent (USPTO) data based on International Patent Classification (IPC). *Scientometrics*, *98*(3), 1583-1599.

Leydesdorff, L., & Vaughan, L. (2006). Co-occurrence matrices and their applications in information science: Extending ACA to the Web environment. *Journal of the American Society for Information Science and technology*, *57*(12), 1616-1628.

Lin, Y., Liu, Z., Sun, M., Liu, Y., & Zhu, X. (2015). Learning entity and relation embeddings for knowledge graph completion. Twenty-ninth AAAI conference on artificial intelligence,

Luo, J. (2022). Data-driven innovation: What is it? *IEEE Transactions on Engineering Management*.

Luo, J., Sarica, S., & Wood, K. L. (2021). Guiding data-driven design ideation by knowledge distance. *Knowledge-Based Systems*, *218*, 106873.

Mao, Y., & Fung, K. W. (2020). Use of word and graph embedding to measure semantic relatedness between Unified Medical Language System concepts. *Journal of the American Medical Informatics Association*, *27*(10), 1538-1546.

Mohamed, S. K., Nounu, A., & Nováček, V. (2021). Biological applications of knowledge graph embedding models. *Briefings in bioinformatics*, *22*(2), 1679-1693.

Nickel, M., Tresp, V., & Kriegel, H.-P. (2011). A three-way model for collective learning on multi-relational data. Icml,

Sarica, S., Yan, B., & Luo, J. (2020). Data-driven intelligence on innovation and competition: patent overlay network visualization and analytics. *Information Systems Management*, *37*(3), 198-212.

Schoen, A., Villard, L., Laurens, P., Cointet, J.-P., Heimeriks, G., & Alkemade, F. (2012). The network structure of technological developments; Technological distance as a walk on the technology map. Science & Technology Indicators (STI) Conference,

Sun, Z., Deng, Z.-H., Nie, J.-Y., & Tang, J. (2019). Rotate: Knowledge graph embedding by relational rotation in complex space. *arXiv preprint arXiv:1902.10197*.

Teece, D. J., Rumelt, R., Dosi, G., & Winter, S. (1994). Understanding corporate coherence: Theory and evidence. *Journal of economic behavior & organization*, *23*(1), 1-30.

Trouillon, T., Welbl, J., Riedel, S., Gaussier, É., & Bouchard, G. (2016). Complex embeddings for simple link prediction. International conference on machine learning,

Whalen, R., Lungeanu, A., DeChurch, L., & Contractor, N. (2020). Patent similarity data and innovation metrics. *Journal of Empirical Legal Studies*, *17*(3), 615-639.

Yan, B., & Luo, J. (2017a). Filtering patent maps for visualization of diversification paths of inventors and organizations. *Journal of the Association for Information Science and Technology*, *68*(6), 1551-1563.

Yan, B., & Luo, J. (2017b). Measuring technological distance for patent mapping. *Journal of the Association for Information Science and Technology*, *68*(2), 423-437.

Yang, B., Yih, W.-t., He, X., Gao, J., & Deng, L. (2014). Embedding entities and relations for learning and inference in knowledge bases. *arXiv preprint arXiv:1412.6575*.

Yoon, J., & Kim, K. (2012). Detecting signals of new technological opportunities using semantic patent analysis and outlier detection. *Scientometrics*, *90*(2), 445-461.

Zhang, Z., Li, Z., Liu, H., & Xiong, N. N. (2020). Multi-scale dynamic convolutional network for knowledge graph embedding. *IEEE Transactions on Knowledge and Data Engineering*.

---

[i] https://ckg.readthedocs.io/en/latest/INTRO.html

[ii] https://patentsview.org/download/data-download-tables

[iii] We use unique ids for all entities: patents, inventors, assignees, etc., instead of their values. In USPTO, each inventor has a unique inventor_id to distinguish different individuals. For instance, the name “Alan Smith” is quite common and found in many patents. It is disambiguated in the database using various unique identifiers such as “fl:al_ln:smith-130”, “fl:al_ln:smith-133”, “fl:al_ln:smith-116”, etc.

[iv] https://aws-dglke.readthedocs.io/en/latest/

[v] Herein, a domain refers to one of 667 groups or 4-digit IPC codes (e.g., H04N). The home domain refers to the subset of 667 groups where an agent had already obtained a patent, while the rest are target domains.

[vi] https://sutdapac-my.sharepoint.com/:f:/g/personal/guangtong_li_mymail_sutd_edu_sg/Ej03CknRKAZKgKRYXnM8_x0Bz893QCvMKZAu59R2hBeHRg